\documentclass[intlimits,twoside,a4paper]{article}
\usepackage{amsmath,amssymb}
\usepackage[T2A]{fontenc}
\usepackage[cp1251]{inputenc}
\usepackage[eqsecnum]{cmpj}

\issue{2011}{14}{2}{23002}

\doinumber{10.5488/CMP.14.23002}

\title[Fokker-Planck equation for non-Markovian Gaussian systems]%
{Generalized Fokker-Planck equation and its solution for linear
non-Markovian Gaussian systems}

\author[O.Yu. Sliusarenko]{O.Yu. Sliusarenko}
\address{Akhiezer Institute for Theoretical Physics NSC KIPT,
1~Akademichna Str., 61108 Kharkiv, Ukraine
}

\authorcopyright{O.Yu. Sliusarenko, 2011}

\date{Received February 11, 2011}
\begin{document}

\hyphenation{co-lou-red}

\maketitle

\begin{abstract}
In this paper we suggest a consistent approach to derivation of
generalized Fokker-Planck equation (GFPE) for Gaussian
non-Markovian processes with stationary increments. This approach
allows us to construct the probability density function (PDF)
without a need to solve the GFPE. We employ our method to obtain the
GFPE and PDFs for free generalized Brownian motion and the one in
harmonic potential for the case of power-law correlation function of the noise. We prove the fact that the considered systems
may be described with Einstein-Smoluchowski equation at high
viscosity levels and long times. We also compare the results with
those obtained by other authors. At last, we calculate PDF of
thermodynamical work in the stochastic system which consists of a
particle embedded in a harmonic potential moving with constant
velocity, and check the work fluctuation theorem for such a system.

\pacs 05.10.Gg, 52.65.Ff, 02.50.Ey, 05.40.-a
\end{abstract}

\section{Introduction}

The theory of Markovian Brownian motion is successfully used in describing a great variety of experiments and
observations~\cite{chandrasekhar, vankampen, klimontovich,
HaenggiACP}. However, it remains an inapplicable model for the majority of
natural systems where a characteristic time of thermal fluctuations
is comparable to that of a Brownian particle (Gaussian non-Markovian
systems), or where the processes are strongly non-Gaussian (either
Markovian or non-Markovian), all of which results in the fact that the
long-time mean squared displacement does not grow linearly in time
any more, $\left \langle x^2(t)\right \rangle \propto t^\mu$. This
phenomenon is called anomalous diffusion, namely, when $\mu<1$, the
system is said to be subdiffusive, and when $\mu>1$, it is
superdiffusive. Evidently, when $\mu=1$ we have an ordinary Brownian
motion.

There are two paradigmatic models describing anomalous diffusion:
continuous time random walk (CTRW) and fractional Brownian motion
(FBM). The CTRW approach was developed by Montroll and Weiss in
1965~\cite{montroll} for a description of the electric charge
transport in a disordered medium (amorphous
semiconductor)~\cite{montroll2}. This model considers the independent
identically distributed couples of random space-time steps whose
PDFs belong to the domain of attraction of L\'evy stable laws.

Recently, the Markovian L\'evy processes in external fields were studied
by means of Langevin and fractional kinetics technique~\cite{chechkin2, cskLK, chechkin3}.

The second model (FBM) was introduced by Kolmogorov in
1940~\cite{kolmogorov} and later studied by Yaglom~\cite{yaglom}.
The name ``fractional Brownian motion'' belongs to Mandelbrot and
van Ness who suggested a stochastic integral representation of this
process ~\cite{mandelbrot}. FBM is a continuous centered
non-Markovian Gaussian process $X^{(H)}(t)$ with covariance function
\begin{equation}
\left\langle X^{(H)}(t) X^{(H)}(t') \right\rangle = D \left(t^{2H} + t'^{2H} - |t-t'|^{2H}\right),
\end{equation}
or, at large times,
\begin{equation}
\left\langle X^{(H)}(t)^2 \right\rangle = 2D t^{2H},
\end{equation}
where $H$ is Hurst index, $0 < H < 1$, and $D$ is the generalized
diffusion coefficient of the dimension
$\left[D\right]=\rm{cm}^2/\rm{sec}^{2H}$. Previously, the problem
of particle escape from the potential well in the framework of
this model was considered in paper \cite{pre2010} by using the
method of numerical simulation of Langevin equation with
fractional Gaussian noise $Y^{(H)}(t)$. The latter is a
non-Markovian stationary random process which is defined as the
time derivative of FBM and whose autocorrelation function exhibits
a slow decay at infinity as $\left \langle Y^{(H)}(t)Y^{(H)}(0)
\right \rangle \approx 2DH(2H-1)t^{2H-2}$, in contrast to white
Gaussian noise, where $\left \langle Y^{(1/2)}(t) Y^{(1/2)}(t')
\right \rangle = 2D \delta (t-t')$. Power spectral density for
white noise does not depend on frequency, otherwise the noise is
called a coloured noise.

The pioneer work of deriving a differential equation (in essence a Fokker-Planck equation, FPE)  describing ordinary Brownian motion (OBM) was done by Lord Rayleigh~\cite{rayleigh}, within the approach of an absence of external potential and an overdamped discrete motion of a heavy Brownian particle. A more consistent method was developed by Fokker, Smoluchowski and Planck (a detailed historical sketch may be found in~\cite{HaenggiACP}). However, when dealing with the coloured noise case, the above-mentioned approaches are no longer valid.

The most common example of derivation of one-dimensional
Fokker-Planck equation for coloured noise may be found in
paper~\cite{HaenggiACP}; for one-dimensional case it was done
in~\cite{caceres}; for a particular case of a linear oscillator it
was obtained and studied in~\cite{wang}. The multi-dimensional case
was considered in \cite{adelman}.

The theory of generalized Brownian motion (GBM) finds its
applications in many problems of modern physics, biophysics and
astronomy. Indeed, polymers~\cite{doi,granek,farge}, elastic
chains and
membranes~\cite{granek,edwards,freyssingeas,granek2,zilman} and
rough surfaces~\cite{majumdar,krug,toroczkai} can be described by
a continuum elastic model which accounts for their general
stochastic behavior; it was recently shown that the probe particle
in such systems performs FBM \cite{taloni, taloni2}. The
fluctuations of magnetic field in the turbulent plasma of the
Earth's magnetospheric tail turn out to have colour: in the range
of frequencies $\omega \leqslant 10^{ - 2}$ Hz they have the
properties of flicker-noise (their power spectrum is proportional
to $ 1/\omega $). When $\omega$ is about $10^{ - 1}$ Hz, they are
a brown noise with the tendency of ``blackening'' at lower
frequencies, see, e.g., the paper~\cite{zeleniy} and works cited
therein. Moreover, a similar situation is known from experiments
in laboratory plasmas: it was found that the power spectra of the
saturation current, electrostatic potential fluctuations, and the
turbulence-induced flux measured in various plasma
devices~\cite{carreras} have power-law dependencies. At high
frequencies, an asymptotic power fall-off of the fluctuation
spectra with characteristic decay indices close to 2 was denoted;
at intermediate frequencies, the decay indices were about 1,
gaining a weak frequency dependence at the lowest frequencies.

Another important application comes from single-molecule dynamics.
In paper~\cite{xie} it is shown, that the experimental data of
the distance fluctuations between the two components of
fluoresceintyrosine complex can be described within the framework of
the Langevin equation with harmonic potential and coloured source possessing correlation function (CF), which decays as $ t^{ - 0.51 \pm 0.07} $.

Below we present a consistent method of derivation of a multi-dimensional generalized Fokker-Planck equation for linear stochastic systems driven by coloured Gaussian noise paying special attention to the case of coloured Gaussian noise with power-law correlation function.

\section{Basics of the method}

We use the approach to obtaining an ordinary Fokker-Planck equation for linear systems
with delta-correlated noise described in
monograph~\cite{peletminskii} as the basis of the suggested method
for derivation of the generalized Fokker-Planck equation.
The paper continues and extends the previous studies
\cite{ujp} where we considered the GFPE for exponential and
power-law correlation function restricting ourselves only
to space-homogeneous case. Here we study a more general problem for the
power-law correlation function. For the integrity and clarity of
presentation, we give a full description of the method, as well.

First, let us write Langevin equations in multi-dimensional form:
\begin{equation}
\label{eqLEPrim}
\dot \xi _i  =  - a_{ik} \xi _k  + Y_i \left( t \right) + K_i\,.
\end{equation}
Here $\xi _i$ is the generalized coordinate, $a_{ik}$ is the coefficient matrix, $Y_i$ is the external noise,
 $K_i$  is the regular constant force; the dot above $\xi_i$ stands for time derivative.
 Let the initial conditions be $\xi _i \left( {t = 0} \right) = \xi _i \left( 0 \right)$.
 Then, the formal solution of (\ref{eqLEPrim}) is
\begin{equation}
\label{eqLEFormalSol}
\xi_i \left( t; \xi \left( 0 \right) \right)
=  \left( {\re^{ - at}} \right)_{ij} \xi _j \left( 0 \right)
+ \int\limits_0^t \rd\tau \left( {\re^{ - a\left( {t - \tau } \right)} } \right)_{ij}
\left( {Y_j \left( \tau  \right) + K_j } \right),
\end{equation}
where $a \equiv \left\| {a_{ik} } \right\|$ is matrix composed
from the elements of $a_{ik}$\,. The probability density function
(PDF) of the value $\xi _i $ in the moment of time $t$ with the
fixed $\xi \left( 0 \right)$ is evidently a multi-dimensional
Dirac delta-function:
\begin{equation}
f\left( {\xi ,t;\xi \left( 0 \right)} \right)  =  \delta \left( {\xi - \xi \left( {t,\xi \left( 0 \right)} \right)} \right) \equiv \prod\limits_i  \delta \left( {\xi _i  - \xi _i \left( {t,\xi \left( 0 \right)} \right)} \right).
\end{equation}

In case we do not know the exact  $\xi \left( 0 \right)$, but
their initial PDF $ f\left( {\xi \left( 0 \right),0} \right)$, the
PDF at an arbitrary moment of time $t$ will be of the shape:
\begin{equation} \label{eq4}
f\left( {\xi ,t} \right) = \int  \rd\xi \left( 0 \right)f\left( {\xi \left( 0 \right),0} \right)\left\langle {\delta \left( {\xi  - \xi \left( {t,\xi \left( 0 \right)} \right)} \right)} \right\rangle ,
\end{equation}
where $\left\langle {\dots } \right\rangle $ stands for $\int
\rd\tau p_Y \left( \tau  \right)\dots $ and $ p_Y \left( \tau
\right)$ is the PDF of noise. By using the  $n$-dimensional
delta-function representation $\delta \left( \xi  \right) = \left(
{2\pi } \right)^{ - n} \int \rd q \exp\left( {\ri q\xi } \right)$
and taking into account (\ref{eqLEFormalSol}), we have:
\begin{equation}
\label{avgDelta}
\left\langle {\delta \left( {\xi  - \xi \left( {t,\xi \left( 0 \right)} \right)} \right)} \right\rangle
= \left( {2\pi} \right)^{ - n} \displaystyle \int  \rd q\hat G\left( {q,t} \right)
\exp\left[ {\ri q\left( {\xi - \re^{ - at} \xi \left( 0 \right)} \right)} \right],
\end{equation}
where
\begin{equation} \label{eq6}
\hat G\left( {q,t} \right) = \left\langle \exp\Biggl\{ { - \ri
q\int\limits_0^t  \rd\tau \re^{ - a\left( {t - \tau } \right)}
\left( {Y\left( \tau  \right) + K} \right)} \Biggr\}
\right\rangle.
\end{equation}
Here we should remark that we use a matrix notation and the hat indicates that the value is a Fourier image.

Expanding the PDF into Fourier integral
\begin{equation}
f\left( {\xi ,t} \right) = \left( {2\pi } \right)^{ - n} \int \rd q \re^{\ri q\xi } \hat f\left( {q,t} \right),
\end{equation}
we get from (\ref{eq4}) and (\ref{avgDelta}):
\begin{equation}
\label{eq8}
\hat f\left( {q,t} \right) = \hat G\left( {q,t} \right)\hat f\left( {\left( {\re^{ - at} } \right)^T q,0} \right),
\end{equation}
where $\left( {\re^{ - at} } \right)^T$ is a matrix transposed to $\re^{ - at} $.

Hereinafter we assume the random process $Y_i \left( t \right)$ to be a stationary Gaussian process, so that the following relations are true:
\begin{eqnarray}
\label{eq9}
\left\langle {Y_{i_1 } \left( {t_1 } \right)\dots Y_{i_{2n + 1} } \left( {t_{2n + 1} } \right)} \right\rangle  &=& 0,
\nonumber\\ \displaystyle \left\langle {Y_{i_1 } \left( {t_1 } \right)\dots Y_{i_{2n} } \left( {t_{2n} } \right)} \right\rangle &=& \displaystyle \sum  g_{i_1 i_2 } \left( {t_1 - t_2 } \right)\dots g_{i_{2n - 1} i_{2n} } \left( {t_{2n - 1}  - t_{2n} } \right),
\end{eqnarray}
where the summation is executed by all possible pair compositions
of $i_1\,, t_1\, ;i_2\, ,t_2\, ;\dots ;i_{2n}\,, t_{2n}\, .$ The
number of such pairs is $ \left( {2n - 1} \right)!! = 2n!/n!2^n.$
$g_{i_1 i_2 } \left( {t_1  - t_2 } \right)$ is a certain function
of time difference.

Using the exponential function series expansion for (\ref{eq6}) and keeping in mind (\ref{eq9}) we have:
\begin{eqnarray}
\nonumber
\hat G\left( {q,t} \right) &=& \sum\limits_{n = 0}^\infty   \frac{{\left( { - \ri} \right)^{2n} }}{{\left( {2n} \right)!}}\frac{{\left( {2n} \right)!}}{{n!2^n }} 
 \left[ {\int\limits_0^t  \rd t_1 \int\limits_0^t  \rd t_2 q_i \left( {\re^{ - at_1 } } \right)_{ij} q_m \left( {\re^{ - at_1 } } \right)_{ml} g_{jl} \left( {t_1  - t_2 } \right)} \right]^n  \\
  &&{} \times  \exp \left( { - \ri q_i \int\limits_0^t {\rd\tau
\left( {\re^{ - a\tau } } \right)_{ij} K_j } } \right).
\end{eqnarray}
Here and below for simplicity we write $g_{jl} \left( {t_1  - t_2 } \right)$ instead of $g_{j_1 l_2 } \left( {t_1  - t_2 } \right)$.

Introducing the value
\begin{equation}
\label{eqMimPrim} \mathfrak{M}_{im} \left( t \right) =
\frac{1}{2}\int\limits_0^t  \rd t_1 \int\limits_0^t  \rd t_2 \left(
{\re^{ - at_1 } } \right)_{ij} \left( {\re^{ - at_2 } }
\right)_{ml} g_{jl} \left( {t_1  - t_2 } \right),
\end{equation}
we get for $G\left( {q,t} \right)$:
\begin{equation}\label{eqG}
\hat G\left( {q,t} \right) = \exp\left[ { - q_i q_m
\mathfrak{M}_{im} \left( t \right) - \ri q_i \int\limits_0^t {\rd\tau
\left( {\re^{ - a\tau } } \right)_{ij} K_j } } \right].
\end{equation}

\subsection{Fokker-Planck equation}

It may be easily proven that the value $\hat G\left( {q,t} \right)$ obeys the following relation:
\begin{equation}
\label{eq13}
\frac{{\partial \hat G}}{{\partial t}} + q_i a_{ik} \frac{{\partial \hat G}}{{\partial q_k }} + \ri K_i q_i \hat G =  - q_i q_m \mathfrak{D}_{im} \left( t \right)\hat G\left( {q,t} \right),
\end{equation}
where
\begin{equation}
\label{eqDimEq} \mathfrak{D}_{im} \left( t \right) =
\frac{{\rd\mathfrak{M}_{im} }}{{\rd t}} + a_{ik} \mathfrak{M}_{km}
\left( t \right) + a_{mk} \mathfrak{M}_{ik} \left( t \right).
\end{equation}

On the other hand, due to an obvious equality
\begin{equation}
\frac{\partial }{{\partial t}}f\left( {\left( {\re^{ - at} } \right)^T q,0} \right) =  - qa\frac{\partial }{{\partial q}}f\left( {\left( {\re^{ - at} } \right)^T q,0} \right)
\end{equation}
and (\ref{eq8}) we can conclude that the function $f\left( {q,t} \right)$ obeys the same equation as (\ref{eq13}) which after the inverse Fourier transform yields:
\begin{equation}
\label{eqGFPEGeneral} \frac{{\partial f\left( {\xi ,t}
\right)}}{{\partial t}} + K_i \frac{{\partial f\left( {\xi ,t}
\right)}}{{\partial \xi _i }} = \frac{\partial }{{\partial \xi _i
}}\left[ {a_{im} \xi _m f\left( {\xi ,t} \right)} \right] +
\mathfrak{D}_{im} \left( t \right)\frac{{\partial ^2 f\left( {\xi
,t} \right)}}{{\partial \xi_i \partial \xi _m }}\,,
\end{equation}
actually being the generalized Fokker-Planck equation (GFPE).

Now, let us simplify the expressions  (\ref{eqDimEq}). Noticing that
\begin{eqnarray*}
 a_{ik} \left( {\re^{ - at} } \right)_{kj}  &=& \frac{\partial }{{\partial t}}
 \left( {\re^{ - at} } \right)_{ij}\, , \\
 \frac{{\partial g_{jl} \left( {t_1  - t_2 } \right)}}{{\partial t_1 }} &=&  - \frac{{\partial g_{jl} \left( {t_1  - t_2 } \right)}}{{\partial t_2 }}
\end{eqnarray*}
and
\begin{eqnarray*}
 \frac{{\rd\mathfrak{M}_{im} }}{{\rd t}} &=& \frac{1}{2}\int\limits_0^t  \rd t_1 \left( {\re^{ - at_1 } } \right)_{ij} \left( {\re^{ - at} } \right)_{ml} g_{jl} \left( {t_1  - t} \right) + \frac{1}{2}\int\limits_0^t  \rd t_2 \left( {\re^{ - at} } \right)_{ij} \left( {\re^{ - at_2 } } \right)_{ml} g_{jl} \left( {t - t_2 } \right),
\end{eqnarray*}
we arrive at the expression
\begin{equation} \label{eqDimGen}
\mathfrak{D}_{im} \left( t \right) =  \frac{1}{2}\int\limits_0^t
\rd t_1 \left[ {\left( {\re^{ - at_1 } } \right)_{ij} g_{jm} \left(
{\left| {t_1 } \right|} \right)} \right. + \left. {\left( {\re^{ -
at_1 } } \right)_{mj} g_{ij} \left( {\left| {t_1 } \right|} \right)}
\right].
\end{equation}

\subsection{Probability density function} \label{secPDF}

The advantage of the described method is that there is no need in
solving the Fokker-Planck equation to obtain the probability
density function since we have constructed it implicitly at the
stage of the GFPE derivation. Indeed, according to (\ref{eq8}), knowing the Fourier image of the initial PDF $\hat f\left( {q,0}
\right) \equiv \mathfrak{F}\left\{ {f\left( {\xi ,0} \right)}
\right\}$ we can easily get the PDF for an arbitrary moment of time:
\begin{eqnarray}
\label{eqPDF}
f\left( {\xi ,t} \right) &=& \mathfrak{F}^{ - 1} \left\{ {\hat f\left( {q,t} \right)} \right\}
\nonumber\\  &=& \mathfrak{F}^{ - 1} \left\{ {\hat G\left( {q,t} \right)\hat f\left( {\left( {\re^{ - at} } \right)^T q,0} \right)} \right\},
\end{eqnarray}
where $\hat G\left( {q,t} \right)$ is given with (\ref{eqG}).

However, the expressions (\ref{eqMimPrim}) may be rather complicated for direct calculations regarding, e.g., a power-law correlation function $g\left( {t_1  - t_2 } \right)$. By means of integration variables change we get a much more usable expression, because now the internal integral does not contain the correlation function:
\begin{eqnarray}
\label{eqMimSingle} \nonumber \mathfrak{M}_{im} \left( t \right) =
\frac{1}{2}\int\limits_0^t \rd\tau g_{jl} \left( \tau
\right)\int\limits_0^{t - \tau } \rd T\left\{ \left( \re^{ -
a\left( T + \tau \right)}  \right)_{ij} \left( {\re^{ - aT} }
\right)_{ml}   +  \left( {\re^{ - aT} } \right)_{ij} \left( \re^{
- a\left( {T + \tau } \right)}  \right)_{ml}  \right\}.
\end{eqnarray}

\section{Generalized Brownian motion}
Let us now apply the derived formulae to the specific stochastic system:  the generalization of the classical Brownian motion with the external random force is a stationary Gaussian noise with long memory effects.

\subsection{Free generalized Brownian motion. Spatially homogenous case}

First, we investigate a simple system described with the following Langevin equations:
\begin{eqnarray}
\label{eqLEsm}
\nonumber \frac{{\rd x}}{{\rd t}} &=& v, \\
\frac{{\rd v}}{{\rd t}} &=&  - \gamma v + Y\left( t \right),
\end{eqnarray}
where $x(t)$ is particles coordinate, $\gamma$ is friction constant, $Y(t)$ is Gaussian external noise with $\left \langle Y(t)Y(t') \right \rangle \equiv g(t-t')$. The coefficient matrix $a_{im}$ is
\[
a = \left[ {\begin{array}{*{20}c}
 0 & { - 1}  \\
   {0} & \gamma   \\
\end{array}} \right],
\]
and $ g_{ij} \left( {t_1  - t_2 } \right) = \delta _{i2} \delta _{j2} g\left( {t_1  - t_2 } \right).$ The solution of the homogenous system (\ref{eqLEsm}) yields:
\begin{eqnarray}
x(t) &=& x_0 + \frac{1-\re^{-t \gamma }  }{\gamma } v_0\,, \nonumber\\
 v(t) &=& v_0 \re^{-t \gamma }.
\end{eqnarray}
Comparing these expressions with (\ref{eqLEFormalSol}) we get
\begin{equation}
 \re^{ - at}  =  \left[ {\begin{array}{*{20}c}
   {1} & {\displaystyle \frac{1-\re^{-t \gamma }}{\gamma }}  \\
   {0} & {\displaystyle \re^{-t \gamma }}  \\
\end{array}} \right].
\end{equation}

In what follows we restrict ourselves to the power-law noise correlation function of the form
\begin{equation}\label{eqPLCF}
g (\tau) = \frac{c}{|\tau|^\beta \Gamma(1-\beta)}\,,
\end{equation}
with $0<\beta<1$, which is actually the asymptotics
of the CF for fractional Gaussian noise. Note, that at $\beta \rightarrow 1$, we get the delta-function limit $g(\tau) \rightarrow c\delta(\tau)$~\cite{gelfand}.

Now, we can write out the exact values for the coefficients
$\mathfrak{D}_{ij}$ and $\mathfrak{M}_{ij}$\,, see
equations~(\ref{eqDimGen}) and~(\ref{eqMimSingle}), respectively:
\begin{eqnarray}
\mathfrak{D}_{11}(t) &=& 0,\\[1ex]
\mathfrak{D}_{12}(t) &=& \mathfrak{D}_{21}(t) = \frac{c t^{1-\beta } }
{2 \gamma \Gamma (2-\beta)} \left[1 + (1 - \beta ) \mathrm{E}_{\beta }(t \gamma )\right] -\frac{1}{2} c \gamma ^{\beta -2}\,,\\[1ex]
\mathfrak{D}_{22}(t) &=& c \gamma ^{\beta -1}-\frac{c t^{1-\beta } \mathrm{E}_{\beta }(t\gamma )}{\Gamma (1-\beta )}\,,\\[1ex]
\mathfrak{M}_{11}(t) &=& -\frac{c \left(1-\re^{-t \gamma }\right) t^{1-\beta}}{\gamma ^3 \Gamma (2-\beta )}+\frac{c t^{2-\beta}}{\gamma ^2 \Gamma (3-\beta )}-\frac{1}{2} c \gamma^{\beta -4}
\nonumber\\[1ex] &&{}+ \frac{c \left(-\re^{-2 t \gamma }+2 \re^{-t \gamma}\right)\mathrm{M}(1-\beta ,2-\beta ,t \gamma ) t^{1-\beta }}{2 \gamma^3 \Gamma (2-\beta )}+\frac{c \mathrm{E}_\beta (t \gamma) t^{1-\beta }}{2 \gamma ^3 \Gamma (1-\beta )}\,, \\[1ex]
\mathfrak{M}_{12}(t) &=& \frac{c \left(1-\re^{-t \gamma }\right)t^{1-\beta }}{2\gamma ^2 \Gamma (2-\beta )} -\frac{c \left(-\re^{-2 t \gamma }+\re^{-t \gamma }\right)t^{1-\beta } \mathrm{M}(1-\beta ,2-\beta ,t \gamma )}{2 \gamma^2 \Gamma (2-\beta )}\,,\\[1ex]
\mathfrak{M}_{22}(t) &=& \frac{1}{2} c \gamma ^{\beta -2} -\frac{c t^{1-\beta } \mathrm{E}_\beta (t \gamma )}{2\gamma  \Gamma (1-\beta )}-\frac{c \re^{-2 t \gamma } t^{1-\beta } \mathrm{M}(1-\beta ,2-\beta,t \gamma )}{2 \gamma  \Gamma (2-\beta )}\,.
\end{eqnarray}
Here $\mathrm{E}_\beta(t)$ is an integral exponential function
\[
\mathrm{E}_\beta(t) = \int_{1}^{\infty} {\rd p \frac{\mathrm{e}^{-t p}}{p^\beta}}
\]
and $\mathrm{M}(a,b,t)$ is Kummer's confluent hypergeometric function:
\[
\mathrm{M}(a,b,t) = \frac{\Gamma(b)}{\Gamma(a)\Gamma(b-a)} \int _0^1 {\rd u \mathrm{e}^{t u} u^{a-1} (1-u)^{b-a-1}},
\]
(see, e.g.~\cite{abramovits}).

The generalized Fokker-Planck equation (\ref{eqGFPEGeneral}) for this case yields:
\begin{equation}
\label{eqGFPEhom} \frac{\partial f(v,t)}{\partial t} =
\frac{\partial }{\partial v}\left[ \gamma v f(v,t)\right] +
D_{22}(t) \frac{\partial^2 f(v,t)}{\partial v^2}\,.
\end{equation}
When $\gamma t \gg 1$, the latter expression takes the form
\begin{equation}
\label{eqGFPEhomt} \frac{\partial f(v,t)}{\partial t} =
\frac{\partial }{\partial v}\left[ \gamma v f(v,t)\right] +
\frac{c}{\gamma} \left(\gamma^\beta - \frac{\mathrm{e}^{-t \gamma}
t^{-\beta}}{ \Gamma(1-\beta)} \right) \frac{\partial^2
f(v,t)}{\partial v^2}\,.
\end{equation}

According to the procedure described in section~\ref{secPDF}, the PDF $f(v,t)$ with the initial condition $f(v,0)=n \delta (v-v_0)$ reads
\begin{equation}
f(v,t) = \frac{n}{2 \sqrt{\pi \sigma }}\, \exp\left[ -\frac{\left(v-\re^{-t \gamma } v_0\right)^2}{4 \sigma } \right],
\end{equation}
where
\begin{equation}
\sigma = \frac{1}{2} c \gamma ^{\beta -2} -\frac{c t^{1-\beta } \mathrm{E}_\beta (t \gamma )}{2 \gamma  \Gamma (1-\beta )} -\frac{c \re^
{-2 t \gamma } t^{1-\beta } \mathrm{M}(1-\beta ,2-\beta,t \gamma )}{2 \gamma  \Gamma (2-\beta )}\,.
\end{equation}

\subsection{Free generalized Brownian motion. Spatially inhomogenous case}

Now we examine the same system but with inhomogenous initial condition $f(x,v,0)=n \delta (x-x_0) \delta (v-v_0)$. All the values for $\mathfrak{D}_{ij}(t)$ and $\mathfrak{M}_{ij}(t)$ clearly, remain the same as in the previous section,
but the GFPE and the PDF do change:
\begin{equation}
\frac{\partial f}{\partial t} = -v \frac{\partial f}{\partial x}
+ \gamma \frac{\partial (v f)}{\partial v}
+ 2\mathfrak{D}_{12} \frac{\partial^2 f}{\partial x \partial v}
+ \mathfrak{D}_{22} \frac{\partial^2 f}{\partial v^2}\,.
\end{equation}

Again we construct the solution with the procedure explained in section~\ref{secPDF}:
\begin{equation}
f(x,v,t|\omega=0)=\frac{n}{4 \pi  \sqrt{\sigma}} \exp
\left[-\frac{1}{4 \sigma} {\mathfrak{M}_{11}(t) \left(p\frac{
   \mathfrak{M}_{12}(t)}{\mathfrak{M}_{11}(t)}-v_0 \mathrm{e}^{-t \gamma
   }+v\right)^2}-\frac{p^2}{4\mathfrak{M}_{11}(t)}\right],
\end{equation}
where
\begin{equation}
p=\frac{v_0}{\gamma } \left(1-\re^{-t \gamma }\right) -x+x_0
\end{equation}
and
\begin{equation}
\sigma = \mathfrak{M}_{11}(t)\mathfrak{M}_{22}(t)-\mathfrak{M}_{12}(t)^2.
\end{equation}

\subsection{Generalized Brownian motion of linear oscillator}
\label{gbmlof}

Here we study the most general system, though restricting ourselves to the case of a
harmonic potential $U\left( x \right) = \omega ^2 x^2 /2$ . The pair of Langevin equations now have the following form:
\begin{eqnarray}
\label{eqLE}
\nonumber \frac{{\rd x}}{{\rd t}} &=& v, \\
 \frac{{\rd v}}{{\rd t}} &=&  - \gamma v - \omega ^2 x + Y\left( t \right),
\end{eqnarray}
where $x(t)$ is particles coordinate, $\gamma$ is friction constant, $\omega$ is frequency of the linear oscillator, $Y(t)$ is the external noise. The coefficient matrix introduced in equation~(\ref{eqLEPrim}) is
\[
 a = \left[ {\begin{array}{*{20}c}
   0 & { - 1}  \\
   {\omega ^2 } & \gamma   \\
\end{array}} \right].
\]
Again, $ g_{ij} \left( {t_1  - t_2 } \right) = \delta
_{i2} \delta _{j2} g\left( {t_1  - t_2 } \right).$ The solution of
the homogenous system (\ref{eqLE}) yields
\begin{eqnarray} \label{eqLESolution}
\nonumber v\left( t \right)  &=& A_1 \re^{ - \gamma t/2} \re^{\Omega t/2}  + A_2 \re^{ - \gamma t/2} \re^{ - \Omega t/2} , \\
 x\left( t \right) &=&  - \frac{1}{{\omega ^2 }}\left[ {\dot v\left( t \right) + \gamma v\left( t \right)} \right]
 \nonumber\\ &=&  - \frac{1}{{2\omega ^2 }}\Big[ A_1 \re^{ - \gamma t/2} \re^{\Omega t/2} \left( {\gamma  + \Omega } \right)
+   {A_2 \re^{ - \gamma t/2} \re^{ - \Omega t/2} \left( {\gamma  - \Omega } \right)} \Big],
\end{eqnarray}
where $A_1$ and $A_2$ are constants depending on the initial
conditions and here we introduce the value $\Omega  \equiv + \sqrt
{\gamma ^2  - 4\omega ^2 }$. Assigning $x\left( 0 \right) = x_0 $
and $v\left( 0 \right) = v_0$\,, we get
\begin{eqnarray}
\nonumber A_1  &=&  - \frac{{2\omega ^2 x_0  + v_0 \left( {\gamma  -
\Omega } \right)}}{{2\Omega }}\,, \\
A_2  &=& \frac{{2\omega ^2 x_0  + v_0 \left( {\gamma  + \Omega }
\right)}}{{2\Omega }}\,.
\end{eqnarray}

Now, substituting the latter expressions into equations (\ref{eqLESolution}) and comparing the result with the formal solution (\ref{eqLEFormalSol}) without the integral term (since we are looking for the solution of the homogenous system), we find
\begin{equation}
\mathrm{e}^{ - at}  = \re^{ - \gamma t/2} \left[ {\begin{array}{*{20}c}
   {\cosh\left( {\displaystyle  \frac{{\Omega t}}{2}} \right) + \displaystyle \frac{\gamma }{\Omega }\sinh\left( {\displaystyle \frac{{\Omega t}}{2}} \right)} & {\displaystyle \frac{2}{\Omega }\sinh\left( {\displaystyle  \frac{{\Omega t}}{2}} \right)}  \\
   { - \displaystyle \frac{{2\omega ^2 }}{\Omega }\sinh\left( {\displaystyle \frac{{\Omega t}}{2}} \right)} & {\cosh\left( {\displaystyle \frac{{\Omega t}}{2}} \right) - \displaystyle \frac{\gamma }{\Omega }\sinh\left( {\displaystyle \frac{{\Omega t}}{2}} \right)}  \\
\end{array}} \right].
\end{equation}

The final step before proceeding to the GFPE and the PDF evaluation is to obtain the exact expressions for $\mathfrak{M}_{ij} \left( t \right)$ and the generalized diffusion coefficients $\mathfrak{D}_{ij} \left( t \right)$ $(i,j = 1,2)$ for our power-law correlation function (\ref{eqPLCF}).

A straightforward integration of equation (\ref{eqMimSingle}) with (\ref{eqPLCF}) gives
\begin{eqnarray}
\label{eqM11}
\nonumber \mathfrak{M}_{11}(t) &=& -\frac{c \re^{-2 a_{p} t} \left(a_{m} \gamma -2 \re^{t \Omega } \omega ^2\right) \mathrm{M}(1-\beta ,2-\beta ,a_{p} t) t^{1-\beta }}{2 \gamma  \omega ^2 \Omega ^2 \Gamma (2-\beta )}\\[1ex]
&&{}-\frac{c \re^{-t \gamma } \left(a_{p} \re^{t \Omega } \gamma -2 \omega ^2\right) \mathrm{M}(1-\beta ,2-\beta ,a_{m} t) t^{1-\beta }}{2 \gamma  \omega ^2 \Omega ^2 \Gamma (2-\beta)}\nonumber\\[1ex]
 &&{}+\frac{a_{m} c \mathrm{E}_\beta(a_{p} t) t^{1-\beta }}{2 \gamma  \omega ^2 \Omega  \Gamma (1-\beta )} - \frac{a_{p} c \mathrm{E}_\beta(a_{m} t) t^{1-\beta }}{2 \gamma  \omega ^2 \Omega  \Gamma (1-\beta)} + \frac{\left(a_{p}^2 a_{m}^{\beta }-a_{p}^{\beta } a_{m}^2\right) c}{2 a_{p} a_{m} \gamma  \omega^2 \Omega }\,, \\[1ex] \label{eqM12}
\mathfrak{M}_{12}(t) &=& -\frac{c \re^{-t (2 a_{p}+\gamma)}}{2 \Omega ^2 \Gamma (2-\beta )}  t^{1-\beta } \Big[ \left(\re^{2a_{p} t}-\re^{t \gamma }\right) \mathrm{M}(1-\beta ,2-\beta, a_{p} t)
\nonumber \\[1ex] &&{}+  \left(\re^{2 a_{p} t}-\re^{t (\gamma +2 \Omega )}\right) \mathrm{M}(1-\beta    ,2-\beta ,a_{m} t)\Big], \\[1ex]
\label{eqM22}
\nonumber \mathfrak{M}_{22}(t) &=& -\frac{a_{p} c \re^{-2 a_{p} t} \left(\gamma -2 a_{m} \re^{t \Omega }\right) \mathrm{M}(1-\beta ,2-\beta ,a_{p} t) t^{1-\beta }}{2 \gamma  \Omega ^2 \Gamma (2-\beta )}\\[1ex]
&&{}-\frac{a_{m} c \re^{t (\Omega -2 a_{p})} \left(\re^{t \Omega } \gamma -2 a_{p}\right) \mathrm{M}(1-\beta ,2-\beta ,a_{m} t) t^{1-\beta }}{2 \gamma  \Omega ^2 \Gamma (2-\beta    )}
\nonumber \\[1ex]&&{}-\frac{a_{p} c \mathrm{E}_\beta(a_{p} t) t^{1-\beta }}
{2 \gamma  \Omega  \Gamma (1-\beta)}+\frac{a_{m} c \mathrm{E}_\beta(a_{m} t)
t^{1-\beta }}{2 \gamma  \Omega  \Gamma (1-\beta )}
+\frac{\left(a_{p}^{\beta}-a_{m}^{\beta }\right) c}{2 \gamma  \Omega }\,,
\end{eqnarray}
where $\Omega = \sqrt{\gamma^2-4\omega^2}$, $a_{p} = (\gamma +
\Omega)/2$, $a_{m} = (\gamma - \Omega)/2$.

According to equation (\ref{eqGFPEGeneral}), the GFPE for such a system reads
\begin{eqnarray}
\label{eqGFPE}
\nonumber \frac{{\partial f\left( {x,v,t} \right)}}{{\partial t}} &=&
\frac{\partial}{\partial x} \left[-v f(x,v,t)\right] + \frac{\partial}{\partial v}\left[ (\omega^2 x+\gamma v)f(x,v,t) \right] \\[1ex]
&&{}+ \mathfrak{D}_{11}\frac{\partial^2 f}{\partial x^2} +
\left(\mathfrak{D}_{12} + \mathfrak{D}_{21}\right)\frac{\partial^2
f}{\partial x \partial v} + \mathfrak{D}_{22}\frac{\partial^2
f}{\partial v^2}\,,
\end{eqnarray}
where
\begin{eqnarray}
\mathfrak{D}_{11} &=& 0,\\[1ex]
\mathfrak{D}_{12}&=&\mathfrak{D}_{21}=\frac{a_m a_p c
\left[\mathrm{E}_{\beta }(a_p t)-\mathrm{E}_{\beta }(a_m t)\right]
t^{1-\beta }}{2 \omega ^2 \Omega \Gamma (1-\beta )}
+\frac{c \left[a_m^{\beta }(\gamma +\Omega ) - a_p^{\beta } (\gamma - \Omega)\right]}
{4 \omega ^2 \Omega }\,,\\[1ex]
\mathfrak{D}_{22} &=& \frac{c \left[a_m \mathrm{E}_{\beta }(a_m
t)-a_p \mathrm{E}_{\beta }(a_p t)\right] t^{1-\beta }}{\Omega
\Gamma (1-\beta )}+\frac{c \left(a_p^{\beta }-a_m^{\beta
}\right)}{\Omega }\,.
\end{eqnarray}

At this stage we may compare these diffusion coefficients to that
obtained in the paper by Wang and Masoliver~\cite{wang}. We
consider only the case of the external driving noise (see
{section~3.2} of the mentioned paper). To establish a connection
with our GFPE and equation~(W29) (here the letter ``W'' indicates
the reference to the equation from the paper~\cite{wang}), let us
substitute equations~(W54),~(W55) and~(W14) into~(W35). Now we
see, that $\psi (t) \equiv 2\mathfrak{D}_{12}(t)$, $\phi (t)
\equiv \mathfrak{D}_{22}(t)$, i.e. we get a complete coincidence
between our GFPE~(\ref{eqGFPE}) and Wang's GFPE~(W29).

The PDF is evaluated directly through relations (\ref{eqPDF}) and
(\ref{eqG}) with $K_j \equiv 0$:
\begin{eqnarray}
\label{eqPDFxvt}
f(x,v,t)&=&\frac{n}{4 \pi  \sqrt{\mathfrak{M}_{11}(t) \mathfrak{M}_{22}(t)-\mathfrak{M}_{12}(t)^2}}
\exp \Bigg( \frac{\re^{- \gamma t}}{4 \Omega ^2
\left(\mathfrak{M}_{11}(t)
\mathfrak{M}_{22}(t)-\mathfrak{M}_{12}(t)^2\right)} \nonumber\\
&&{}\times  \bigg \{ \mathfrak{M}_{11}(t) \left[\re^{\gamma t/2} v
\Omega - v_0 \cosh \left(\Omega t/2\right) \Omega +\left(2 x_0
\omega ^2+ v_0 \gamma \right) \sinh \left(\Omega
t/2\right)\right]^2 \nonumber\\ &&{}+ \mathfrak{M}_{22}(t)
\left[-\re^{\gamma t/2} x \Omega +x_0 \cosh \left(\Omega
t/2\right) \Omega +(2 v_0+x_0 \gamma ) \sinh \left(\Omega
t/2\right)\right]^2 \nonumber \\&&{}+ 2 \mathfrak{M}_{12}(t)
\left[\re^{\gamma t/2} v \Omega -v_0 \cosh    \left(\Omega
t/2\right) \Omega +\left(2 x_0 \omega ^2+v_0 \gamma \right) \sinh
\left(\Omega t/2\right)\right] \nonumber\\ &&{}\times
\left[-\re^{\gamma t/2} x \Omega +x_0 \cosh \left(\Omega t/2
\right) \Omega +(2 v_0+x_0 \gamma ) \sinh \left(\Omega t/2\right)
\right] \bigg \} \Bigg),
\end{eqnarray}
where $\mathfrak{M}_{ij}$ are given with equations~(\ref{eqM11}--\ref{eqM22}).

\section{Transition to Einstein-Smoluchowski equation}

Now let us prove that the system considered in
section~\ref{gbmlof} may be described with Einstein-Smoluchowski
equation at high viscosity levels and at long times.

When $\omega /\gamma \ll 1$, we can neglect the time derivative of velocity and, therefore, the pair of Langevin equations~(\ref{eqLE}) transforms into a single overdamped Langevin equation:
\begin{equation}\label{eqLEover}
\frac{\rd x}{\rd t} = -\frac{\omega^2}{\gamma} x(t) + \frac{1}{\gamma} Y(t).
\end{equation}

The GFPE for such a system, according to equations~(\ref{eqLEover}) and (\ref{eqGFPEGeneral}) has the following form:
\begin{equation}
\frac{\partial f(x,t)}{\partial t} = \frac{\partial }{\partial x} \left( \frac{\omega^2}{\gamma} x f(x,t) \right) + \mathfrak{D}(t)\frac{\partial^2 f(x,t)}{\partial x^2}
\end{equation}
with
\begin{equation}
\mathfrak{D}(t) = \frac{c \omega ^{2\beta - 2} }{\gamma^ {\beta + 1} } \left(1-\frac{\Gamma \left(1-\beta ,t \omega ^2 / \gamma \right)}{\Gamma (1-\beta )}\right).
\end{equation}

Executing the same calculations as in the previous section, for the PDF we unfold:
\begin{equation}
\label{eqPDFoverdamped}
\rho(x,t) = \frac{n}{2 \sqrt{\pi \mathfrak{M}(t)}} \exp \left[ {-\frac{\left(x-x_0 \re^{-t    \omega ^2/\gamma}\right)^2}{4 \mathfrak{M}(t)}} \right],
\end{equation}
where
\begin{eqnarray}
\label{eqMoverdamped}
\mathfrak{M}(t) &=& \frac{1}{2} c \gamma ^{-\beta } \omega ^{2 \beta -4} %
 -\frac{c t^{1-\beta } }{2 \gamma  \omega ^2 \Gamma (1-\beta )} \left[\mathrm{E}_\beta\left(\frac{t \omega ^2}{\gamma}\right) + \frac{\re^{-2 t \omega ^2/\gamma } }{1-\beta } \mathrm{M}\left(1-\beta ,2 -\beta,\frac{t \omega ^2}{\gamma }\right)\right].\qquad
\end{eqnarray}

Similar results were obtained by M.~C\'aceres in~\cite{caceres}
for a stationary case [see equations (2.14) with (2.17) of the
mentioned paper].

Expanding the coefficient~(\ref{eqMoverdamped}) into a series at large $(\gamma t)$'s and
substituting it to the PDF~(\ref{eqPDFoverdamped}) yield:
\begin{eqnarray}
\label{eqPDFoverappr}
\rho(x,t) & \approx & \frac{n \gamma ^{\beta /2} \omega ^{2-\beta}}{\sqrt{2 \pi c }} \left( 1 + \frac{t^{-\beta } \gamma ^{\beta } \omega ^{-2 \beta } \re^{-t \omega ^2 /\gamma }}{\Gamma (1-\beta )}\right)
\nonumber\\ &&{} \times  \exp \left[ - x^2 \left( \frac{\gamma ^{\beta } \omega ^{4-2 \beta }}{2 c} + \frac{t^{-\beta } \gamma ^{2 \beta } \omega ^{4-4 \beta } \re^{-t \omega ^2 /\gamma }}{c \Gamma (1-\beta )}\right)+\frac{x x_0 \gamma ^{\beta } \omega ^{4-2 \beta } \re^{-t \omega ^2 /\gamma }}{c} \right].
\end{eqnarray}

Now we return to the PDF for the most general case~(\ref{eqPDFxvt}) and also expand it into a series at $\omega/\gamma \ll 1\ll \gamma t$:
\begin{eqnarray}
f (x,v,t) &\approx & A \exp \left[ -v^2 \left(-\frac{t^{-\beta } \gamma ^{2-2 \beta } \re^{-\omega ^2 t/\gamma }}{c \Gamma (1-\beta )}+\frac{t^{-\beta } \re^{-t \gamma } \omega ^{2-2 \beta }}{c \Gamma (1-\beta )}+\frac{\gamma ^{-\beta } \left(\gamma ^2-3 \omega ^2\right)}{2 c}\right) \right.\nonumber\\
\nonumber &&{}- x^2 \left(-\frac{t^{-2 \beta } \re^{-t \gamma } \gamma ^{\beta } \omega ^{4-4 \beta }}{c \Gamma (1-\beta )^2}+\frac{t^{-\beta } \gamma ^{2 \beta } \omega ^{4-4 \beta } \re^{-\omega ^2 t/\gamma}}{c \Gamma (1-\beta )}+\frac{\omega ^{4-2 \beta } \left(\gamma ^{\beta }-\gamma ^{-\beta } \omega ^{2 \beta }\right)}{2 c}\right) \\
\nonumber &&{}- v \left( \frac{ \omega ^2 \gamma ^{-\beta } \re^{-\omega ^2 t/\gamma} ({v_0}+{x_0} \gamma )}{c}-\frac{{v_0} \re^{-t \gamma } \gamma ^{2-\beta }}{c}\right) \\
\nonumber &&{}- x \left(\frac{{v_0} \re^{-t \gamma } \gamma ^{\beta -1} \omega ^{4-2 \beta   }}{c}-\frac{\gamma ^{\beta -1} \omega ^{4-2 \beta } \re^{-\omega ^2 t/\gamma} ({v_0}+{x_0}   \gamma )}{c}\right) \\
\nonumber &&{}- {v_0}^2 \left(\frac{\gamma ^{\beta -2} \omega ^{4-2 \beta } \re^{-2\omega ^2 t/\gamma}}{2 c}-\frac{\re^{-t \gamma } \gamma ^{\beta -2} \omega ^{4-2 \beta }}{c}\right) \\
\nonumber &&{}- {x_0}^2 \left(\frac{\gamma ^{\beta } \omega ^{4-2 \beta } \re^{-2\omega ^2 t/\gamma}}{2 c}-\frac{\re^{-t \gamma } \gamma ^{\beta -2} \omega ^{6-2 \beta }}{c}\right) \\
 &&{}- \left. {v_0} {x_0} \left(\frac{\gamma ^{\beta -1} \omega ^{4-2 \beta }    \re^{-2\omega ^2 t/\gamma}}{c}-\frac{\re^{-t \gamma } \gamma ^{\beta -1} \omega ^{4-2 \beta    }}{c}\right) \right],
\\
A &\approx& \frac{2n \gamma  \omega ^{2-\beta }}{4 \pi c} \left(1+\frac{\omega ^{-2 \beta } {\gamma^\beta }t^{-\beta } \re^{-{\omega ^2 t}/\gamma }}{\Gamma (1-\beta )}-\frac{\re^{-t   \gamma } (t \omega )^{-2 \beta }}{\Gamma (1-\beta )^2}\right).
\end{eqnarray}
Then, integrating it by $v$ in the range of $(-\infty; \infty)$
and neglecting the terms of the higher magnitude of smallness than
$\exp\left\{- \omega ^2 t/\gamma \right\}$ we get:
\begin{equation}
\tilde{f} (x,t) \propto  \exp \left[ - x^2 \left( \frac{\gamma^{\beta } \omega ^{4-2 \beta }}{2 c} + \frac{t^{-\beta } \gamma ^{2 \beta } \omega ^{4-4 \beta } \re^{- \omega ^2 t/\gamma }}{c \Gamma (1-\beta )}\right)+\frac{x x_0 \gamma ^{\beta } \omega ^{4-2 \beta } \re^{-t\omega ^2 /\gamma }}{c}   \right],
\end{equation}
which fully corresponds to the PDF~(\ref{eqPDFoverappr}), and,
therefore, proves the fact that the considered system at large
times and strong friction may be described with
Einstein-Smoluchowski equation.

\section{GFPE for overdamped harmonic oscillator with constant drift}

As a final application example of the presented technique, let us
study the PDF of the thermodynamical work $w$ in the stochastic
system which consists of a particle inside a harmonic potential
moving with constant velocity $v_*$\,, $U=(k/2)[x-X(t)]^2$,
$X(t)=v_*t$, $x\left(t\right)$ is the particle's coordinate. Our
aim is to get the transient fluctuation relation for such a
system, which will demonstrate large-deviation symmetry properties
in the PDF, and compare it to the classical case.

The work $w$ is defined as follows:
\begin{equation}
w(t)=\int \rd X \frac{\partial U}{\partial X} = \int_0^t \rd t' \frac{\rd X}{\rd t'} \frac{\partial U}{\partial X} = -kv_* \int_0^t \rd t'(x-v_*t').
\end{equation}

Introducing $y\left(t\right) = x(t) -v_{*} t$, for the overdamped Langevin equation and the equation
for the thermodynamical work $w\left(t\right)$ we have:
\begin{eqnarray}
\frac{\rd y}{\rd t} &=&  -\frac{1}{\tau } y\left(t\right)+Y \left(t\right)-v_{*}\,,
\nonumber\\ \frac{\rd w}{\rd t} &=& -kv_{*} y\left(t\right),
\end{eqnarray}
where $\tau =m\gamma/ k $. Alternatively, if we consider
the plane $\left(y,w\right)$, the coefficient matrix  $a$ of the
system will be
\begin{equation}
a=\left[\begin{array}{cc} {1/\tau } & {0} \\ {kv_{*} } & {0} \end{array}\right].
\end{equation}

Since $y\left(t\right)=y_{0} \exp(-t/\tau) $, $y_{0} =x_{0} $ is the initial position of the particle,
\[
w\left(t\right)=w_0+y_{0} kv_{*} \tau \left(\re^{-t/\tau } -1\right),
\]
Then the evolution matrix
\begin{equation}
\re^{-at} =\left[\begin{array}{cc} {\re^{-t/\tau } } & {0} \\ {kv_{*} \tau \left(\re^{-t/\tau } -1\right)} & {1} \end{array}\right].
\end{equation}

For the diffusion coefficients $\mathfrak{D}    _{ij} \left(t\right)$ we have:
\begin{eqnarray}
\mathfrak{D}_{11}(t) &=& c \gamma ^2 \tau ^{1-\beta } \left(1-\frac{\Gamma \left(1-\beta    ,t/\tau \right)}{\Gamma (1-\beta )}\right),\\
\mathfrak{D}_{12}(t) &=& \frac{c k v_* \gamma ^2 t^{-\beta }
\left(t/\tau \left[(\beta -1)    \mathrm{E}_{\beta }\left(t/\tau \right)-1\right]
+  \Gamma (2-\beta ) \left( t /\tau \right)^{\beta } \right)}{2 \Gamma (2-\beta )}\,,\\[1ex]
\mathfrak{D}_{22}(t) &=& 0.
\end{eqnarray}

The generalized Fokker-Planck equation in this case will have the
form:
\begin{equation}
\frac{\partial f\left(y,w,t\right)}{\partial t}
=\left(\frac{y}{\tau } +v_{*} \right)\frac{\partial f}{\partial y}
+kv_{*} y\frac{\partial f}{\partial w} +\mathfrak{D}_{11} \frac{\partial^2 f}{\partial y^2}
 +2\mathfrak{D}_{12} \left(t\right)\frac{\partial ^{2} f}{\partial y\partial w}\,.
\end{equation}

Now, considering an initial condition $f\left(y,w,0\right)=n\delta
\left(y-y_0 \right)\delta \left(w-w_0\right)$, when $y_0=0$,
$w_0=0$ we get for the PDF:
\begin{equation}
f\left(w,t\right)= \frac{n}{2 \sqrt{\pi }
\sqrt{\mathfrak{M}_{22}(t)}} \exp \left\{ -\frac{\left[k v_*^2
\tau ^2 \left(t/\tau + \re^{-t/\tau}-1\right)+w\right]^2}{4
\mathfrak{M}_{22}(t)} \right\},
\end{equation}
where
\begin{eqnarray}
\mathfrak{M}_{22} &=& \frac{c m^2 v_*^2 \gamma ^2 t^{2-\beta }}{2 \Gamma (3-\beta )} \left\{ 2-\re^{-t/\tau} \rm{M}\left[1,3-\beta ,\frac{t}{\tau }\right] \right.
\nonumber\\& &{}- \left. \re^{-t/\tau} \left(2-\re^{-t/\tau}\right) \rm{M}\left[2-\beta ,3-\beta
   ,\frac{t}{\tau }\right] \right\},
\end{eqnarray}
which fully corresponds to the results obtained in~\cite{chekla}.

After the relaxation stage has passed, at $t\gg \tau$ we find for the transient fluctuation relation:
\begin{equation}
\frac{f\left(w,t\right)}{f\left(-w,t\right)} =\exp \left[\frac{\Gamma (3-\beta ) w t^{\beta -1}}{c m \gamma} \right].
\end{equation}
Thus, the fluctuation relation for the system subjected to a
coloured noise with the slowly decaying power-law correlation
function differs from that for ordinary Brownian motion. As we
stated above, the classical case limit is revealed at $\beta
\rightarrow 1$.

\section{Conclusions}
In this paper we suggested a consistent method for derivation of the
generalized Fokker-Planck equation for linear multidimensional
Gaussian non-Markovian systems. Taking the case of the Gaussian
systems with slowly decaying power-law correlations, we obtained the
following results:
\begin{itemize}
\item Firstly, we constructed the solution of generalized Fokker-Planck equation, the probability density function, without solving it directly.
\item We derived generalized Fokker-Planck equation for free motion and constructed the probability density function for spatially homogeneous and inhomogeneous cases.
\item For the case of the motion in a harmonic potential, the generalized Fokker-Planck equation and the probability density function were also obtained, and the results were compared to those of the other authors.
\item We show the equivalence in description of generalized Brownian motion in a harmonic potential with generalized Fokker-Planck equation and generalized Einstein-Smoluchowski equation at high viscosity levels and at long times.
\item Finally, we investigated the probability density function for thermodynamical work in the
stochastic system which consists of a particle inside a uniformly moving harmonic
potential underlining strong differences in transient fluctuation relations for the generalized Brownian motion and the ordinary Brownian motion cases.
\end{itemize}

\section*{Acknowledgements}
O.~S. would like to thank A.V.~Chechkin for the problem setting
and for the discussion of the results, S.V.~Peletminskii and
Yu.V.~Slyusarenko for helpful comments on the paper.

\newpage

\ukrainianpart

\title{Узагальнене рівняння Фокера-Планка та його розв'язок для лінійних немарківських Ґаусових систем}
\author{О.Ю. Слюсаренко}
\address{Інститут теоретичної фізики ім. О.І. Ахієзера ННЦ ХФТІ, Україна, 61108 Харків, вул. Академічна, 1
}

\makeukrtitle

\begin{abstract}
\tolerance=3000%
У цій роботі ми пропонуємо послідовний підхід до виводу
узагальненого рівняння Фокера-Планка (УРФП) для Ґаусових
немарківських процесів із стаціонарними прирощеннями. Цей підхід
дозволяє побудувати функцію розподілу (ФР) процесу без потреби
безпосередньо розв'язувати УРФП. Ми застосовуємо цей метод для
знаходження УРФП та ФР для вільного узагальненого броунівського руху та
узагальненого броунівського руху в потенціалі для випадку степеневої
кореляційної функ\-ції шуму. Ми доводимо, що
розглянуті системи можуть описуватися у рамках рівняння
Ейнштейна-Смолуховського за умов сильної в'язкості та
великих часів. Також ми порівнюємо результати із отриманими іншими
авторами. Нарешті, ми обчислюємо ФР термодинамічної роботи у
стохастичній сис\-те\-мі, що складається з частинки у гармонічному
потенціалі, який рухається з постійною швидкістю, та перевіряємо
флуктуаційну теорему для роботи у такій системі.

\keywords рівняння Фокера-Планка, Ґаусова система,
немарківська система, термодинамічна робота, перехідне флуктуаційне
співвідношення

\end{abstract}

\end{document}